# The low temperature heat capacity of solutions of methane isotopes in fullerite $C_{60}$. Isotope effects

M.I. Bagatskii[1], V.G. Manzhelii[1], V.V. Sumarokov[1], A.V. Dolbin[1], M.S. Barabashko[1], B. Sundqvist[2]

[1] B. Verkin Institute for Low Temperature Physics & Engineering NASU, Kharkov 61103, Ukraine

[2] Department of Physics, Umea University, SE - 901 87 Umea, Sweden

Electronic address: bagatskii@ilt.kharkov.ua

Abstract.

The heat capacity $C(T)$ of the interstitial solid solution $(CH_4)_{0.4}C_{60}$ has been investigated in the temperature interval 1.4-120 K. The contribution ($C_{CH4}$) of $CH_4$ molecules to the heat capacity $C$ has been separated. The contributions ($C_{CH4}$) and ($C_{CD4}$) of $CH_4$ and $CD_4$ molecules to the heat capacity of the solutions $(CH_4)_{0.40}C_{60}$ and $(CD_4)_{0.40}C_{60}$ have been compared.

It is found that above 80 K the character of the rotational motion of $CH_4$ and $CD_4$ molecules changes from libration to hindered rotation. In the interval 14-35 K the heat capacities $C_{CH4}(T)$ and $C_{CD4}(T)$ are satisfactorily described by contributions of the translational and libration vibrations, as well as the tunnel rotation for the equilibrium distribution of the nuclear spin species. The isotope effect in $C_{CH4}(T)$ and $C_{CD4}(T)$ is due, mainly, the difference in the frequencies of local tranlational and libration vibrations of molecules $CH_4$ and $CD_4$.

The contribution of the tunnel rotation of the $CH_4$ and $CD_4$ molecules to the heat capacities $C_{CH4}(T)$ and $C_{CD4}(T)$ is dominant below 8 K. The isotopic effect is caused by the difference between both the conversion rates and the rotational spectra of the nuclear spin species of $CH_4$ and $CD_4$ molecules. The conversion rate of $CH_4$ molecules is several times lower than that of $CD_4$ ones.

Weak features observed in the curves $C_{CH4}(T)$ and $C_{CD4}(T)$ near 6 K and 8 K, respectively, are most likely a manifestation of first-order phase transitions in the orientational glasses of these solutions.

**Introduction**

Investigations of the physical properties of solid solutions $(CH_4)_nC_{60}$ and $(CD_4)_nC_{60}$ can provide valuable information about the isotopic effects and their role in the low temperature dynamics of isolated spherical rotators in the octahedral cavities of the low temperature phase Pa3 of the fullerite lattice.

The difference between the molecular masses of methane and deuteromethane is relatively small ($M_{CD4}/M_{CH4} = 1.25$) but the substances differ significantly in moments of inertia ($I_{CD4}/I_{CH4} = 2$), total nuclear spins of the A -, T- and E- species of $CH_4$ and $CD_4$ molecules ($S_{CH4} = 2, 1, 0$ and $S_{CD4} = 4, 2, 0$, respectively) and magnetic moments of the light atoms ($\mu_H/\mu_D=3.268$). These differences can cause isotopic effects in the heat capacities $C_{CH4}(T)$ and $C_{CD4}(T)$ of the admixed $CH_4$ and $CD_4$ molecules. Substitution of $CD_4$ for $CH_4$ in the solutions has a relatively slight effect on the translational subsystem but changes dramatically the physical properties of the rotational subsystem of the $CH_4$ and $CD_4$ molecules at helium temperatures at which the quantum effects are dominant.

The low temperature dynamics of the solutions $(CH_4)_nC_{60}$ and $(CD_4)_nC_{60}$ was investigated by the methods of neutron [1-6] and X-ray [7,8] diffraction, NMR spectroscopy of $^2H$ [2,3] and $^{13}C$ [9,10] atoms, dilatometry [11,12] and adiabatic calorimetry [13].

Intercalation of fullerite with $CH_4$ and $CD_4$ molecules (and other simple impurities [14-23]) leaves the structures of the high – temperature and low – temperature phases of fullerite unaltered. The lattice parameters of the high-temperature phase of the solutions $(CH_4)_{0.91}C_{60}$ and $(CD_4)_{0.87} C_{60}$ increase at $T\approx 300$ K by 0.026Å and 0.016Å [2,6], respectively, relative to that of pure $C_{60}$ ($a = 14.161$Å [24]). This depresses the molecular interaction and smears the orientational phase transitions to ~241 K in $(CH_4)_{0.91}C_{60}$ and ~ 235 K in $(CD_4)_{0.87}C_{60}$ [2,6] from ~ 260 K in pure $C_{60}$. Below these temperatures the $C_{60}$ molecules have the <111> orientations. There are two orientational configurations of $C_{60}$ molecules in which the orientations of the rotation axes of $C_{60}$ differ by $60^0$. These are symmetrically nonequivalent configurations for $C_{60}$ molecules known as pentagonal (p) and hexagonal (h) [6,25]. The energy difference between the p- and h- configurations is $\Delta E \approx 11$ meV [6,26]. In the orientationally ordered phase the $C_{60}$ molecules execute fast p – h and h – p jumps. As the temperature decreases, the fraction of the p

– configurations increases and the frequency of the jumps attenuates. Below the temperature of glass transition ($T_g \approx$ 90 K) the p – and h – configurations are frozen and the time of reorientation increases to about an hour [6]. The intercalation of impurity atoms (molecules) to the fullerite $C_{60}$ leads to a change in $\Delta E$ and in the p- and h- fractions.

Investigations of the thermal expansion $\alpha(T)$ of the solutions $CH_4 - C_{60}$ and $CD_4 - C_{60}$ show that at $T$=4-6 K the orientational glasses of these solutions undergo first-order phase transitions evidenced as a hysteresis and maxima in the temperature dependences of $\alpha(T)$ [11,12].

The dynamics of the solutions $(CH_4)_{0.91}C_{60}$ and $(CD_4)_{0.87}C_{60}$ was investigated at $T$=1.5-40 K by Kwei et al. [6] using the method of inelastic neutron scattering in the energy range 0.1-100 meV. The energy differences between the lower levels of the rotational spectra of the nuclear spin A-, T- and E- species of $CH_4$ molecules in the octahedral cavities of $C_{60}$ were measured (see Table 1 in Ref. [6]). A nuclear spin conversion of $CH_4$ molecules was revealed and the time $t_{1/2} \approx$2.6 h of the conversion between the ground states of the A- and T- species was found. The authors of Ref. [6] analyzed comprehensively their own results and the data [2,9,10] obtained by other methods. It is found that the impurity molecules have a minor effect on the motion of the molecules $C_{60}$. The $CH_4$ molecules execute rotational vibrations at helium temperatures and a weakly hindered rotation at $T\sim$120 K [6].

The heat capacity of the solution $(CD_4)_{0.40}C_{60}$ in the temperature interval 1.2 – 120 K was measured by Bagatskii et al. [13] The calorimetric results for the rotational – vibrational states of the $CD_4$ molecules in the octahedral cavities in fullerite $C_{60}$ [13] are in good agreement with the inelastic neutron scattering [6] and NMR spectroscopic [2,3] data on the dynamics of the molecular motion in the solutions $(CH_4)_{0.92}C_{60}$ and $(CD_4)_{0.88}C_{60}$.

Our goal was to carry out a calorimetric investigation of the isotopic effects in the low temperature dynamics of admixture molecules in the solid interstitial solutions $(CH_4)_xC_{60}$ and $(CD_4)_xC_{60}$.

**Experiment**

The heat capacity $C(T)$ of the interstitial solid solution $(CH_4)_{0.40}C_{60}$ was investigated under a constant pressure in the interval $T$=1.5-120 K. The difference between the heat capacities at constant pressure and volume is negligible for both fullerite and the solution $(CH_4)_{0.40}C_{60}$ below 120 K [24]. The heat capacity $C(T)$ was obtained by subtracting the addenda heat capacity $C_{ad}(T)$ from the total heat capacity $C_{ad+sol}(T)$. The temperature dependence of the $C_{ad}(T)$ (the heat capacity of an "empty" calorimeter with the Apiezon vacuum grease) was measured in a separate experiment [27,28]. The measurements were made by the heat pulse technique using an adiabatic vacuum calorimeter [27]. The variation of the sample temperature was $\Delta T_i = T_{i+1} - T_i$, where $T_i$ and $T_{i+1}$ are the temperatures of the sample before switching on and after switching off heating. It was 0.2-10% of $T_i$ in a single measurement run. The heat capacity corresponds to the average temperature $T=(T_{i+1}+T_i)/2$. The heating time $t_h$ was 1-10 minutes. The time required to measure one value of heat capacity was 0.1-0.4 hour. This was dictated mainly by the time $t_e$ of equalizing the temperature in the sample after switching off heating. The characteristic time $t_m$ of a single heat capacity measurement was taken to be $t_m=t_h+t_e$. The temperature of the calorimeter was measured with a "CERNOX" thermometer [27,28].

The sample was a cylinder ~8 mm high and 10 mm in diameter. The samples were prepared from a high-purity (99.99%) $C_{60}$ powder (SES, USA) with a grain size of about 0.1 mm. The $C_{60}$ powder was saturated with methane/deuteromethane under similar conditions: $P \approx$200 MPa and $T$=575ºC for 36 hours. The procedure was performed at the Australian Science and Technology Organization, Australia. The $CH_4$ or $CD_4$ molecules filled about 50% of the octahedral cavities in the samples according to Thermal gravimetric analysis (TGA). The $CH_4$ and $CD_4$ – saturated $C_{60}$ powders were compacted at Umea University, Sweden. The technology of sample preparation and estimation of the methane (deuteromethane) concentration are described in [29,30]. The obtained results of the thermal expansion of the $CH_4$-$C_{60}$ and $CD_4$-$C_{60}$ samples are presented in Refs. [11,12]. After completing the dilatometric measurements, an X-ray diffraction study of the $CH_4$-$C_{60}$ sample was performed in the vicinity of the orientational phase transition, $T$=140-320 K [8].

The concentration n ≈ 40 mol% of admixture methane molecules was measured in a separate experiment with a low temperature vacuum desorption gas analyzer [31] after completing the measurement of heat capacity. The design and operation of the gas analyzer is detailed in [31].

The mass of the sample was $m$ = 874.23 ± 0,05 mg. The sample weighing and mounting in the calorimeter along with hermetization of the vacuum chamber took several hours. Then the vacuum chamber of the calorimeter was washed several times with pure nitrogen gas and the sample was held in the dynamic vacuum at room temperature for ~ 14 hours. The residual $N_2$ pressure in the vacuum chamber was up to several mTorr. The calorimeter was cooled through wires without using He as a heat exchanging gas. The cooling from room temperature to ≈ 5 K took about eight hours. The cooling of the calorimeter from 5 K to about 1.4 K and the reaching a steady–state temperature rate of $10^{-3}$ – $10^{-4}$ K/min took about 10 hours. Several series of measurement were performed. The measurement results were only slightly dependent on the temperature prehistory of the sample.

The measurement error in the heat capacity of $(CH_4)_{0.40}$ $(C60)$ was ± 15 % at $T$=1.5 K, ± 4% at $T$= 2 K and ± 2% at $T \geq 4$ K.

**Results and discussion**

The measured heat capacity $C(T)$ of the solid solution $(CH_4)_{0.40}C_{60}$ normalized to unit mole of $C_{60}$ is shown in Fig. 1 (a: $T$=1.4–120 K; b: $T$=1.4 – 4 K). The heat capacity $C_f$ of pure fullerite $C_{60}$ added on the figure to comparison has been measured previously in the same calorimeter [27,28]. The heat capacity of the solution increases with temperature in the whole of temperature region. The ratio $C/C_f$ is about 9 at 1.5 K, 5 at 2 K, 2 at 4 K, 1.5 at 20 K and decreases to 1.3 at 116 K.

Proceeding from the analysis of inelastic neutron scattering data for the $(CH_4)_{0.92}C_{60}$ and $(CD_4)_{0.88}C_{60}$ solutions a conclusion was drawn [6] that intercalation of fullerite with both $CH_4$ and $CD_4$ molecules caused only a slight decrease in the frequencies of the translational and libration lattice modes of $C_{60}$. Therefore, the increase in $C(T)$ against $C_f(T)$ is mainly due to the rotational and translational motion of the $CH_4$ molecules in the octahedral cavities of the $C_{60}$ lattice.

The dependence $C(T)$ is analyzed assuming that doping of $C_{60}$ with $CH_4$ molecules has a negligible small effect on the lattice vibrations of the $C_{60}$ molecules [6]. The heat capacity component $C_{CH4}$ was obtained by subtracting $C_f(T)$ of pure $C_{60}$ [13,27] from the heat capacity $C$ of the solution $(CH_4)_{0.40}C_{60}$ ($C_{CH4}(T) = C(T) - C_f(T)$).

The temperature dependence $C_{CH4}(T)$ normalized to the unit mole of $CH_4$ in the solution is shown in Fig. 2 (a: $T$ = 1.4 – 120 K; b: $T$ = 1.4 –18 K) along with the calculated heat capacity curves determined by local translational ($C_{tr}$, curve 2) and libration ($C_{lib}$, curve 3 ) vibrations and by tunnel rotation of the $CH_4$ molecules in the potential field of the octahedral cavities of the $C_{60}$ lattice for the equilibrium ($C_{rot,eq}$, curve 4) and "frozen" high–temperature ($C_{rot,eq}$, curve 5 in Fig. 2b) distributions of the nuclear spin species of the $CH_4$ molecules. The total heat capacity $C_{calc} = C_{tr} + C_{lib} + C_{rot,eq}$ is described by curve 1. The $C_{tr}$ and $C_{lib}$ were calculated within the Einstein model. The $C_{tr}$ was calculated using the characteristic Einstein temperature $\Theta_{tr}$ = 126.5 K found from inelastic neutron scattering data [6]. The characteristic Einstein temperature $\Theta_{lib}$ = 67 K used to calculate $C_{lib}$ was found from the condition of the best fit to $C_{CH4}(T)$. The calculation of $C_{rot,eq}$ and $C_{rot,high}$ is described below. It is seen in Fig. 2a that experimental $C_{CH4}(T)$ and calculated $C_{calc}$ agree well in the region $T$ = 8 – 35 K.

As the temperature rises above 35 K, the curve $C_{CH4}(T)$ goes upwards and reaches a maximum at ≈ 90 K. On a further increase in the temperature the curve $C_{CH4}(T)$ descends. The fall of $C_{CH4}(T)$ at $T > 90$ K occurs because the rotational motion of the CH$_4$ molecules changes from libration to hindered rotation.

At $T = 35 - 90$ K the calculated heat capacity $C_{calc}$ is lower than $C_{CH4}(T)$. As for the solution (CD$_4$)$_{0.40}$(C$_{60}$), we assume that the discrepancy may be due to "additional" rotational excitations of the CH$_4$ molecules in the region of a phase transition from a glass state to an orientationally ordered phase. According to Ref. [6], the motion of CH$_4$ molecules has a weak effect on the translational and rotational lattice vibrations of the C$_{60}$ molecules in the solution; on the contrary, the motion of the C$_{60}$ molecules affects appreciably the local translational and rotational vibrations of the CH$_4$ molecules. The formations of an orientation glass in the solution (CH$_4$)$_{0.40}$C$_{60}$ changes the character of the rotational motion of the C$_{60}$ molecules, which induces the "additional" rotational excitations of the CH$_4$ molecules.

The contribution of the tunnel rotation of the CH$_4$ molecules to $C_{CH4}(T)$ dominates below 8 K (see Fig. 2b). At $T<8$ K the heat capacity $C_{CH4}(T)$ is determined by the low-lying energy levels of the rotational spectra of the nuclear spin A - , T- and E - species of the CH$_4$ molecules and by the correlation between the characteristic time $\tau_{CH4}$ of CH$_4$ conversion and the time $t_m$ of one measurement run [32,33]. The low – energy parts of the rotational spectra of the A -, T- and E - species of CH$_4$ are as for CD$_4$ dependent on the symmetry and value of the crystal field. The rotators CH$_4$ and CD$_4$ are therefore an effective probe of the crystal environment in the octahedral cavities of the fullerite C$_{60}$. The A- species has the lowest – energy state. Therefore, under the equilibrium condition at $T = 0$ K all the CH$_4$ molecules are in this state. At $T > 0$ K the equilibrium distribution of the A-, T- and E- species is reached through state conversion.

The energy differences between the low – lying energy levels of the A-, T- and E- species of the CH$_4$ molecule in the octahedral cavity of the lattice in the solution (CH$_4$)$_{0.92}$(C$_{60}$) were found by the method of inelastic neutron scattering [6] (see Table 1 in Ref. [6]). It is found [6] that (i) the rotational spectra are close in the p- and h- configurations; (ii) the relaxation time of the occupancy of rotational energy levels above the ground state of the A- and T- species of CH$_4$ molecules is rather short; (iii) the time of conversion between the ground states of the A- and T- species of CH$_4$ molecules at $T \approx 4$ K is $t_{1/2} \approx 2.6$ h. A two-parameter model of a crystal potential field $V(w) = B_{CH4}[\beta_4 V_4(w) + \beta_6 V_6(w)]$ in the octahedral cavity of C$_{60}$ was proposed. Here the $B_{CH4} = 7.538$ K is the rotational constant of the CH$_4$ molecule, $V_4(w)$, $V_6(w)$ are basis functions with symmetry $\bar{A}_1 A_1$ and $\beta_4$, $\beta_6$ are the dimensionless parameters. Proceeding from the available experimental data and assuming $\beta_4 = 2.1856\beta_6$ Kwei et. al. calculated the energies of some low-lying levels in the rotational spectra of the A-, T- and E-species of CH$_4$ molecules as a function of the potential field parameters ($\beta_4$, $\beta_6$) (see Fig.1 in Ref.[6]).

Figure 3 illustrates our estimates (based on the data of Table 1 in [6]) of the low-energy parts of the rotational spectra of CH$_4$ molecules in the octahedral cavity of fullerite C$_{60}$. The $C_{rot,eq}(T)$ (Fig.2a, curve 4) was calculated for the case of "fast" conversion ($\tau_{CH4} \ll t_m$) when the concentration distribution of the nuclear spin species of CH$_4$ at the test temperature can be considered as being in equilibrium at any moments. This occurs when the conversion is completed within the effective time $t_m$ of a single measurement. Under such a condition the heat capacity can be calculated using a unified rotational energy spectrum including all species levels.

The $C_{rot,high}(T)$ was calculated (see Fig. 2a, curve 5) for the case when there is no convertion ($\tau_{CH4} \gg t_{ex} \gg t_m$). Under this condition the distribution of species remains constant during the measurement time $t_m$ (several weeks) in the whole temperature region and is equal to the high-temperature "frozen" distribution $x_A:x_B:x_E = 5:9:2$. Therefore, $C_{rot,high}(T) = (5/16)C_A(T) + (9/16)C_T(T) + (2/16)C_E(T)$, where the molar heat capacities $C_{A,rot}(T)$, $C_{T,rot}(T)$, $C_{E,rot}(T)$ depend only on the transitions inside each species.

It is seen in Fig. 2b that the experimental heat capacity $C_{CH4}(T)$ is inconsistent with the calculated curves $C_{rot,eq}(T)$ and $C_{rot,high}(T)$. At $\tau_{CH4} \sim t_m$ the heat capacity $C_{CH4}(T)$ depends on the number of CH$_4$ molecules which convert during the time $t_m$ [32]. The characteristic time $\tau_{CH4}(T)$ of the conversion in CH$_4$ molecules can be described as

$$\tau_{CH4}(T) = -t_m / \ln(1 - K'), \qquad (1)$$

where $K'= C_{CH4}(T)/C_{rot,eq}(T)$ is the fraction of CH$_4$ molecules in the equilibrium distribution that converted during the time $t_m$.

The time of cooling of the calorimeter from 5 K to 1.4 K and achieving a steady-state rate of temperature was about 10 hours. During this period the CH$_4$ species came to a near-equilibrium distribution over the sample. Most of the CH$_4$ molecules were in the ground state of the A – species. In our experiment $\tau_{CH4} \approx t_m$ and $C_{CH4}(T)$ is determined by the number of CH$_4$ molecules which converted during the effective time $t_m$ of a single heat capacity measurement [32].

Above 8 K $C_{CH4}(T)$ corresponds to the equilibrium distribution of CH$_4$ species. The relatively fast drop of the theoretical $C_{rot,eq}(T)$ values in the region $T = 8$-$12$ K suggests that the higher-lying energy levels of the molecules CH$_4$ (than for those shown in Fig.3) are ignored in calculation of $C_{rot,eq}(T)$.

We estimated $K'=C_{CH4}(T)/C_{rot,eq}(T)$ and the average experimental times $t_m$ at $T = 2$-$7.5$ K. The temperature dependence of the characteristic conversion time $\tau_{CH4}(T)$ of the CH$_4$ molecules in the octahedral cavities of C$_{60}$ was calculated using the $K'(T)$ and $t_m(T)$ values according Eq(1). The dependences $\tau_{CH4}(T)$ and $\tau_{CD4}(T)$ [13] for the solutions (CH$_4$)$_{0.40}$C$_{60}$ and (CD$_4$)$_{0.40}$C$_{60}$ are illustrated in Fig.4. The curves are in good qualitative agreement. The $\tau_{CH4}(T)$ and $\tau_{CD4}(T)$ decrease with rising temperature. At $T = 4$ K $\tau_{CH4}(T)$ is about 5.5 times lower than $t_{1/2}$ and about 4.7 times higher than $\tau_{CD4}(T)$ [13].

The isotopic effects in $C_{CH4}(T)$ and $C_{CD4}(T)$ are illustrated in Fig. 5 (a: $T = 1.2$-$120$ K; b: $T = 1.2$-$18$ K). In the interval T = 12-40 K the isotopic effect is mainly due to the frequency difference between the local translational and libration vibrations of CH$_4$ and CD$_4$ molecules. The Einstein temperatures of the translational ($\theta_{CH4,tr}$, $\theta_{CD4,tr}$) and libration ($\theta_{CH4,lib}$, $\theta_{CD4,lib}$) vibrations of CH$_4$ and CD$_4$ molecules are given in Table 1.

Table 1. The Einstein temperatures of translational ($\theta_{CH4,tr}$, $\theta_{CD4,tr}$) and librational ($\theta_{CH4,lib}$, $\theta_{CD4,lib}$) vibrations of CH$_4$ and CD$_4$ [13] molecules in octahedral cavities of fullerite C$_{60}$.

|                 | $\theta_{tr}$, K | $\theta_{lib}$, K |
|-----------------|------------------|-------------------|
| CH$_4$          | 126.5            | 67                |
| CD$_4$          | 112.6            | 51                |

In the region of the orientational phase transition from a glass state to an orientationally ordered phase of fullerite C$_{60}$ ($\Delta T$=40-100 K) $C_{CD4}(T)$ and $C_{CH4}(T)$ are influenced by "additional" excitations of CH$_4$ and CD$_4$ molecules (Fig.5a). These excitations appear because the motion of the C$_{60}$ molecules in the lattice fullerite and the motion of the impurity molecules are coupled.

At $T$=80-120 K the curve $C_{CH4}(T)$ is higher than the curve $C_{CD4}(T)$ (Fig.5a). This may be due to an uncertainty in the CH$_4$ and CD$_4$ concentrations. The behavior of $C_{CD4}(T)$ and $C_{CH4}(T)$ correlates with the effects of the impurity molecules CH$_4$ and CD$_4$ on change of the lattice parameter of C$_{60}$. At $T \sim 300$ K the lattice parameter of the solutions (CH$_4$)$_{0.91}$C$_{60}$ and (CD$_4$)$_{0.87}$C$_{60}$ increases by 0,026Å and 0,016Å [6], respectively, relative to that of pure C$_{60}$ (a=14,161Å) [24].

The $C_{CH4}(T)$ and $C_{CD4}(T)$ exhibit a complex isotopic effect below 14 K (Fig.5b) when the contributions of the translational and libration modes decrease exponentially and quantum regularities scale up to a macroscopic level. The dependences $C_{CH4}(T)$ below 8 K and $C_{CD4}(T)$ below 6 K are determined by the tunnel rotation of the CH$_4$ and CD$_4$ molecules. The isotope effect is due to differences between both the rates of conversion (see Fig. 4) and rotational spectra of nuclear spin modifications of molecules CH$_4$ and CD$_4$ (see Fig. 3 in this article and Fig. 4 in Ref. [13]). The rotational constant of CH$_4$ molecules is twice as large as that of CD$_4$ molecules. Therefore, the distance between the rotational tunnel energy levels of the $CH_4$ species is also twice that in the case of $CD_4$. The differences in the level degeneracy are due to the different total nuclear spin species of the molecules. In the case of an equilibrium distribution of $CH_4$ and $CD_4$ species the isotopic effect in $C_{CH4}(T)$ and $C_{CD4}(T)$ makes itself evident in the differences between the curves describing $C_{rot,eq}$ (Fig. 5b, solid curve for $CH_4$ and dashed curve for $CD_4$). Besides, below 5 K the conversion rate of $CD_4$ molecules is several times higher than that of $CH_4$ molecules (Fig.4). The conversion of molecules deuteromethane have been first found in the solid solutions $(CD_4)_xKr$ [35,36]. The behavior of the experimental temperature dependences $\tau_{CH4}(T)$ and $\tau_{CD4}(T)$ [13], as in cases of solid solutions $(CH_4)_xKr$ [32,33], $(CD_4)_xKr$ [34,35], agrees qualitatively with the theoretical data on conversion in multiatomic molecules [36-38].

In contrast to diatomic molecules [37-40], other mechanisms of conversion are dominant in multiatomic molecules at low temperatures. These are hybrid [41] and "quantum relaxation" [42,43] mechanisms. According to the theory of the hybrid conversion mechanism [41], the highest conversion rate is determined by the intramolecular magnetic interaction of protons and the intermolecular noncentral interaction between the nearest molecules in the lattice. The intramolecular interaction mixes the nuclear spin states, while the intermolecular noncentral interaction induces transitions between the rotational states transferring the conversion energy to the lattice. The hybrid mechanism is dominant at low temperatures [41]. The noncentral interaction between $CD_4$ and $C_{60}$ molecules is more than in case of $CH_4$ and $C_{60}$ molecules. This is because $CD_4$ molecules have smaller amplitudes of zero-point orientational vibrations and a larger effective electric octupole moment in the ground state. Therefore, the probability of conversion energy transfer from the $CD_4$ molecules to the $C_{60}$ lattice is higher than in the case of $CH_4$ molecules. Hence, $CD_4$ molecules have a higher conversion rate.

The contribution of the "quantum relaxation" mechanism of conversion increases with temperature [42,43]. The conversion rate is determined by the intramolecular magnetic interaction of photons and the tunnel exchange of the states of the species having equal energy levels (no phonons are involved). The distance between the lowest-lying rotational levels of the $CD_4$ molecules is half as large as that in case of $CH_4$ molecules (see Fig. 3 in this work and Fig. 4 in Ref. [13]). Therefore, at equal temperatures below 5 K the conversion rate of $CD_4$ is higher than that of $CH_4$. In the regions $T = 1.3$-$5.5$ K and $T = 1.4$-$8$ K the experimental heat capacities $C_{CD4}(T)$ and $C_{CH4}(T)$ are dependent on the number of the molecules which converted during the effective time $t_m$ of a single run of heat capacity measurement [32]. The heat capacities $C_{CD4}(T)$ above $\approx 5.5$ K and $C_{CH4}(T)$ above $\approx 8$ K correspond to the equilibrium distribution of the molecular species.

It is interesting that at $T = 5$ - $6$ K and $T = 7$ - $8$ K the scatter of the $C_{CD4}(T)$ and $C_{CH4}(T)$ data, respectively, is considerably wider than at the other temperatures. This may be due to the first-order phase transitions in the orientational glasses of the $CH_4$-$C_{60}$ and $CD_4$-$C_{60}$ solutions [11,12].

**Conclusions**

The heat capacity of $CH_4$ –doped fullerite $(CH_4)_{0.4}C_{60}$ has been first investigated in the interval 1.4-120 K. The contribution $C_{CH4}(T)$ of the $CH_4$ molecules isolated in the octahedral cavities of the $C_{60}$ lattice to the heat capacity of the solution as been separated. The contributions of the $CH_4$ and $CD_4$ to the heat capacity of the solutions $(CH_4)_{0.4}C_{60}$ and $(CD_4)_{0.4}C_{60}$ have been compared.

It has been found that above 80 K the rotational motion of the $CH_4$ and $CD_4$ molecules changes from librational vibrations to a hindered rotation. The heat capacities $C_{CH4}(T)$ in the range from 8 to 35 K and $C_{CD4}(T)$ in $T=14$-$35$ K are described well taking into account the contributions of translational and librational vibrations, as well as the tunnel rotation of the $CH_4$ and $CD_4$ molecules in the case of an equilibrium distribution of the nuclear spin species. The isotopic effect is caused mainly by the frequency differences between the local translational and libration vibrations of the $CH_4$ and $CD_4$ molecules.

The characteristic conversion times of the lowest-lying levels of the $CH_4$ species have been estimated at $T<7$ K. The isotopic effect in $C_{CH4}(T)$ and $C_{CD4}(T)$ is induced by the differences between both the conversion rates and the rotational spectra of the nuclear spin species of the $CH_4$ and $CD_4$ molecules.

The weak features observed in the curves $C_{CH4}(T)$ near 6 K and $C_{CD4}(T)$ near 8 K are most likely manifestations of first-order phase transition in the orientational glasses of the solutions [11, 12].

The authors are indebted to A. I. Prokvatilov for a fruitful discussion and to G. E. Gadd, S. Moricca and D. Cassidy for samples preparation.

Fig. 1. The heat capacities of the solid solution $C(T)$ (open circle) $(CH_4)_{0.40}C_{60}$ and pure fullerite $C_f(T)$ [27] (open triangle) normalized to unit mole of $C_{60}$ in the temperature intervals 1.4 – 120 K (a) and 1.4 – 4 K (b).

Fig. 2. The contribution $C_{CH4}(T)$ (open circle) of the $CH_4$ molecules to the heat capacity $C(T)$ of the solution at $T = 1.4 – 120$ K (a) and 1.4 – 18 K (b). The calculated molar heat capacities are determined by local translational ($C_{tr}$, curve 2) and libration ($C_{lib}$, curve 3) vibrations and tunnel rotation of the $CH_4$ molecules for the equilibrium ($C_{rot.eq}$, curve 4) and high temperature ($C_{rot,high}$, curve 5 in Fig. 2b) distributions of the nuclear spin species. The total heat capacity $C_{calc} = C_{tr} + C_{lib} + C_{rot,eq}$ is described by curve 1.

Fig. 3. The low-energy part of the rotational spectrum of free $CH_4$ molecules [34] and for $CH_4$ molecules in the potential field of the octahedral cavities in $C_{60}$ [13] for the A-, T- and E- nuclear spin species. $J$ is the rotational quantum number and $E$ is the energy (the degeneracy of the energy levels are shown in parentheses to the right).

Fig. 4. The characteristic conversion times $\tau_{CH4}(T)$ (circles) and $\tau_{CD4}(T)$ [13] (solid line) of $CH_4$ and $CD_4$ molecules in the octahedral cavities of fullerite.

Fig.5. The isotopic effects in $C_{CH4}(T)$ (open circles) and $C_{CD4}(T)$ [13] (cross) at $T$=1.2-120 K (a) and $T$=1.2-18 K (b). The solid and dashed curves in Fig. 5b are the calculated heat capacities $C_{rot,eq}$ determined by tunnel rotation of $CH_4$ and $CD_4$ molecules, respectively, in the case of the equilibrium distribution of the nuclear spin species.

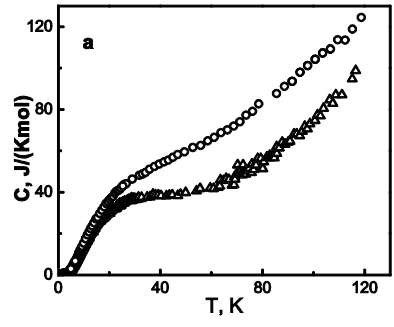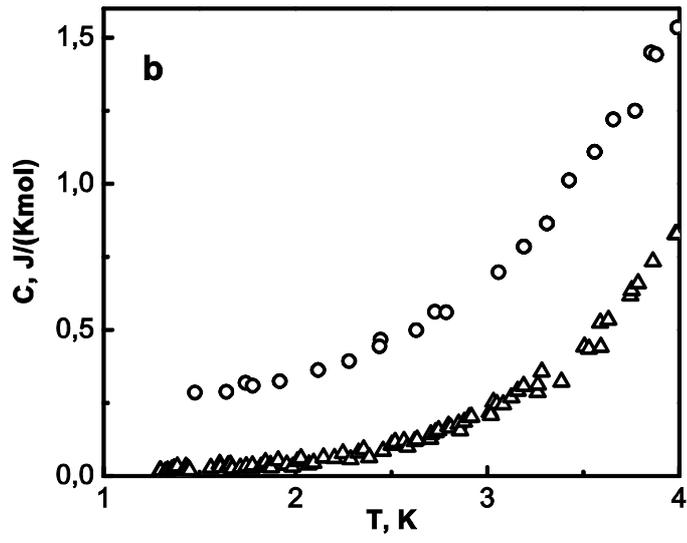

Fig. 1.

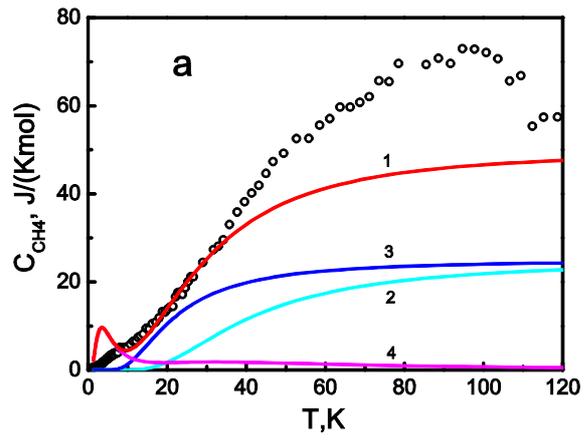

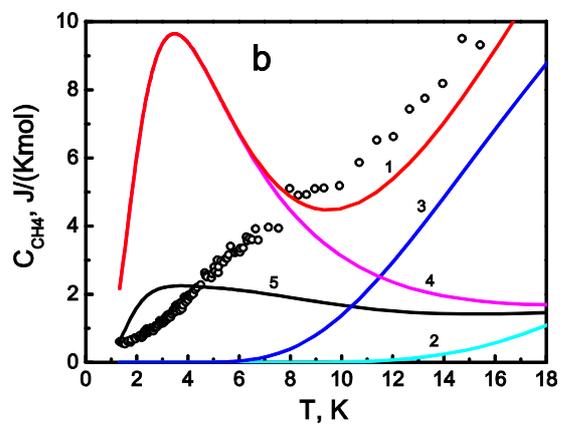

Fig. 2.

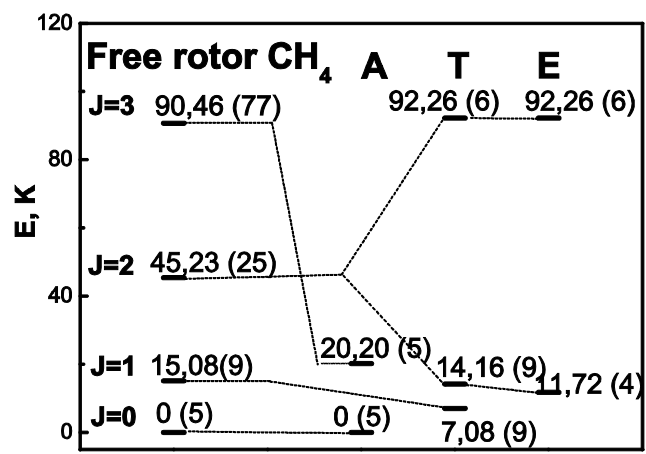

Fig. 3.

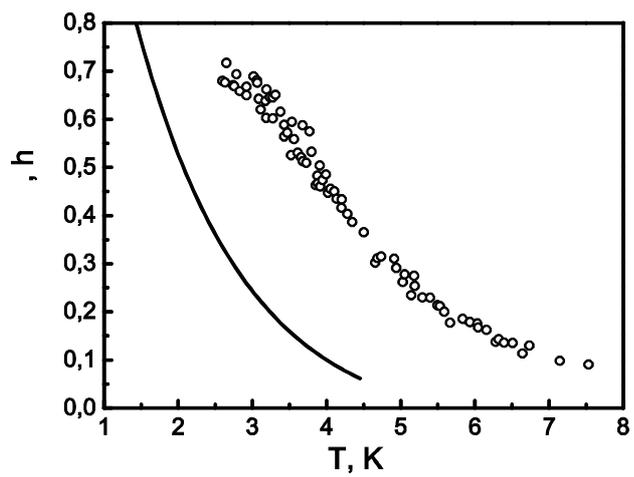

Fig. 4.

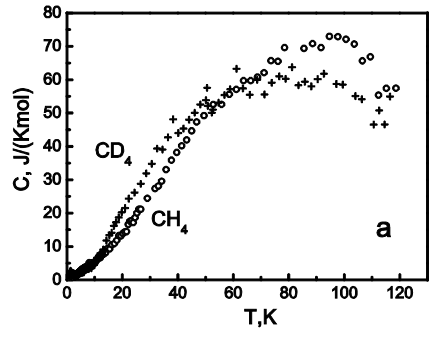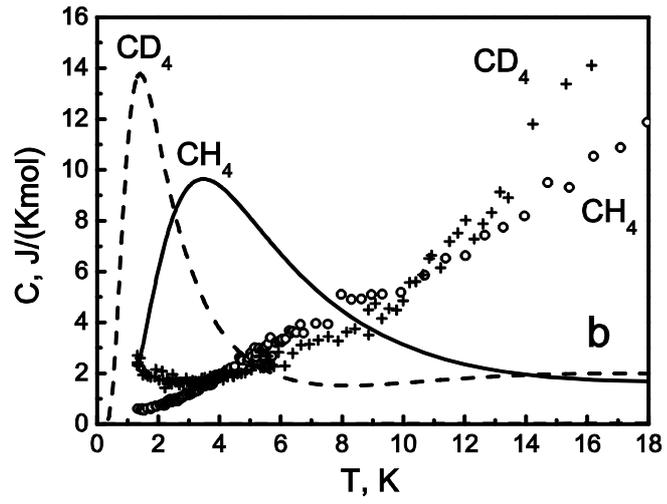

Fig. 5.